\documentclass[seceq,supplement]{ptptex}
\usepackage{graphicx}

\usepackage{wrapft}

\title{Higher-Order Singularities without Glass-Glass Transitions}
\subtitle{An Avoided Glass-Glass Transition}  

\author{Matthias \textsc{Sperl}\footnote{E-mail: matthias.sperl@dlr.de}}

\inst{Institut f\"ur Materialphysik im Weltraum,\\
Deutsches Zentrum f\"ur Luft- und Raumfahrt, \\
51170 K\"oln, Germany}

\abst{

Within the framework of mode-coupling theory, the glass-transition 
scenario is investigated for a system of particles interacting with a 
hard-core repulsion and an additional square-shoulder soft core at larger 
distances. The static structure is calculated from the potential in 
Percus-Yevick approximation. For certain widths of the shoulder, the 
exponent parameter $\lambda$ along the glass-transition line shows a 
double peak. At both peaks, $\lambda$ can reach unity indicating the 
existence of higher-order glass-transition singularities. It is shown that 
these higher-order singularities originate from a line of avoided 
glass-glass transitions.

}

\PTPindex{300, 302}

\begin{document}
\maketitle

\section{Introduction}

In the field of glassy slow dynamics, many experiments and simulations 
have been inspired in recent years by the mode-coupling theory for 
idealized glass transitions (MCT) \cite{Goetze2009}. Within this theory, 
the transition from a liquid to an idealized glass state is described by a 
bifurcation in the solutions of certain polynomials in a variable $f$: 
Upon smooth variations of a control parameter, say, the temperature $T$, 
variable $f$ jumps from $f=0$ to a finite critical value $f=f^c>0$ at 
$T=T_c$ and increases further with the distance from the transition 
$T_c-T$. Variable $f$ is defined by the long-time limit of some 
autocorrelation function $\phi(t)$; states with $f>0$ are identified as 
glass states, while states with $f=0$ shall be called fluid or ergodic.

The mentioned bifurcations originate from the equations of motion for 
density autocorrelation functions $\phi_q(t)$ of a system of $N$ particles 
in a volume $V$ with density $\rho = N/V$. The time-dependent fluctuations 
in the density are used to define the canonical normalized correlation 
function $\phi_q(t) = \langle \rho(t)^*\rho \rangle/\langle \rho^*\rho
\rangle$ as is well-known in liquid-state theory \cite{Hansen1986}. In 
this case, the correlation functions depend on the wave vector modulus $q$ 
of the corresponding Fourier transform in space. Using projection-operator 
techniques, one can derive the following equations of motion

\begin{subequations}\label{eq:eom}
\begin{equation}\label{eq:eom:phi}
\partial_t^2\phi_q(t) + \nu_q\partial_t\phi_q(t) + \Omega_q^2 \phi_q(t)
+  \Omega_q^2\int\,dt'\ m_q(t-t') \partial_{t'}\phi_q(t') = 0\,,
\end{equation}
with characteristic frequencies $\Omega_q$, a white noise $\nu_q$, and the 
memory function
\begin{equation}\label{eq:eom:F}
m_q(t) = {\cal F}[\phi_k(t),V] = 
\frac{1}{2}\int\frac{d^3k}{(2\pi)^3} V_{\vec{q},\vec{k}} 
\phi_k(t)\phi_{|\vec{q}-\vec{k}|}(t)\,,
\end{equation}
where the interaction potential is encoded in the vertex
\begin{equation}\label{eq:eom:V}
V_{\vec{q},\vec{k}} = \rho S_qS_kS_{\vec{q},\vec{k}}\rho
\left[
\vec{q}\vec{k}c_k+\vec{q}(\vec{q}-\vec{k})c_{|\vec{q}-\vec{k}|}
\right]^2/q^4\,,
\end{equation}
through the static structure factor $S_q$ of the fluid and $c_q$ its 
direct correlation function. Both functions can be calculated from the 
interaction potential together with some closure relation that is known
typically only in some approximation \cite{Hansen1986}.
For $t\rightarrow\infty$, one gets an algebraic equation for the 
correlators' long time limits $\phi_q(t)\rightarrow f_q$,
\begin{equation}\label{eq:eom:fq}
\frac{f_q}{1-f_q} = {\cal F}[f_k,V]\,.
\end{equation}
\end{subequations}

It was discovered in 1984 that Eq.~({\ref{eq:eom:fq}) can exhibit 
nontrivial solutions, $f_q > 0$, for microscopic interaction potentials 
and realistic values of the density in the system \cite{Bengtzelius1984}. 
It was shown later that only singularities of type $A_\ell$ can occur in 
Eq.~({\ref{eq:eom:fq}) \cite{Goetze1995b}, and these singularities are 
equivalent to those emerging for the parameter space of roots of 
polynomials upon variation of the coefficients \cite{Arnold1986}. Hence, 
an $A_2$ singularity -- also called \textit{fold} -- signals a double root 
in the solutions like in the equation $x^2 + t = 0$ for $t_c = 0$. For a 
polynomial of high enough order, the variation of a single control 
parameter is sufficient to encounter an $A_2$ singularity. Generically, 
the variation of $\ell - 1$ control parameters is necessary to identify 
singularities of type $A_\ell$. An $A_3$ singularity -- called 
\textit{cusp} -- requires two control parameters to adjust a polynomial to 
a cubic root; an $A_4$ singularity -- the \textit{swallowtail} -- requires 
the variation of three parameters. It was shown in a theorem by Whitney 
that only $A_2$ and $A_3$ are robust singularities, all other 
singularities can be removed by small perturbations of the control 
parameters \cite{Whitney1955,Arnold1986}.

Within MCT, the $A_2$ singularity can be identified with a liquid-glass 
transition if $f$ jumps from $f=0$ to a finite value $f^c$ at the 
transition. Once such a singularity is identified, asymptotic expansions 
of Eq.~(\ref{eq:eom}) can be used to derive the long-time behavior of the 
correlator $\phi(t)$. For the $A_2$ singularity, these asymptotic 
expansions yield two-step relaxation, time-temperature superposition, and 
power-law scaling \cite{Goetze2009}. The unique number characterizing the 
leading terms of the asymptotic expansion for an $A_2$ singularity is the 
exponent parameter $\lambda$ which is between 0.5 and unity. In addition 
to liquid-glass transitions, fold singularities can also describe 
glass-glass transitions: In this case an existing first glass state with 
$f\geq f_1^c$ transforms into a second distinct glass state with $f\geq 
f_2^c > f_1^c$ discontinuously. The endpoint of a line of glass-glass 
transition points is the $A_3$ singularity. An $A_4$ singularity signals 
the emergence of a glass-glass transition line from an otherwise smooth 
surface of liquid-glass transitions. Every $A_\ell$ singularity is 
characterized by a unique number $\mu_\ell \geq 0$ that determines the 
properties of the asymptotic expansions \cite{Goetze2002}. 
$\mu_\ell\rightarrow 0$ signals the emergence of the higher-order 
singularity $A_{\ell+1}$ where in turn $\mu_{\ell+1}$ defines the leading 
terms of the expansion. An $A_2$ singularity's exponent parameter is 
identical to $\mu_2=1-\lambda$. Therefore, a fold gives rise to a cusp 
once $\lambda$ approaches unity.

\section{Glass Transitions, Glass-Glass Transitions, and Higher-Order
Singularities}\label{sec:rev}

In the following, the transition singularities of MCT shall be reviewed 
briefly for hard spheres (sec.~\ref{sec:HSS}), sticky hard spheres 
(sec.~\ref{sec:SHSS}), and the square-well potential (sec.~\ref{sec:SWS}). 
The section~\ref{sec:rev} shall be concluded with the discussion of newly 
discovered transitions in the square-shoulder system (sec.~\ref{sec:SSS}) 
where the static structure factors are evaluated in Roger-Young (RY) 
approximation. In sec.~\ref{sec:SSSPY}, new results shall be given for the 
square-shoulder system with the static structure factors calculated from 
the Percus-Yevick (PY) approximation. The comparison between results from 
RY- and PY-calculations gives insight into the robustness of the predicted 
glass-glass transition phenomena.

\subsection{Glass Transition for the Hard-Sphere System -- a Fold}
\label{sec:HSS}

MCT was first applied to the hard-sphere system (HSS) where a glass 
transition was identified upon varying the control parameter packing 
fraction $\varphi = \pi\rho d^3/6$ for particles of hard-sphere diameter 
$d$ \cite{Bengtzelius1984}. The transition is predicted for a packing 
fraction of $\varphi^c = 0.516$ when using the PY approximation for the 
calculation of the static structure factor $S_q$ \cite{Hansen1986}. The 
hard-sphere interaction can be realized to a very good degree in 
experiments performed in colloidal suspensions. In such experiments the 
glass transition is found around a packing fraction $\varphi^c = 0.58$, 
moreover, the two-step relaxation, scaling laws and several other features 
of the theoretical predictions are confirmed \cite{Megen1991}. From a 
thorough analysis of Eq.~(\ref{eq:eom:V}) one can derive that for the HSS, 
the glass transition is driven by the hard-core repulsion as encoded in 
the principal peak of the static structure factor. The HSS has been 
investigated by asymptotic expansions \cite{Franosch1997} with the 
exponent parameter being around $\lambda = 0.7$ which is close to the 
typical value for many other glass forming substances. The full evolution 
of glassy dynamics over eight orders of magnitude in time for various 
densities has been demonstrated for a tagged particle's mean-squared 
displacement (MSD) \cite{Sperl2005}.

\subsection{Glass-Glass Transition in the Sticky Hard-Sphere System -- a 
Cusp}
\label{sec:SHSS}

For the investigation of an $A_3$~singularity, more than one control 
parameter needs to be varied. Generically the variation of two control 
parameters extends an isolated $A_2$~singularity to a line of 
$A_2$~singularities, so the existence of an $A_3$ singularity is by no 
means certain. However, if such an $A_3$ singularity exists in the given 
parameter plane, it can be found as an endpoint of a line of 
$A_2$~singularities where close to the $A_3$, these $A_2$~singularities 
can be identified as glass-glass transition points. At the $A_3$, the 
distinction between the different glass states, the discontinuity between 
the $f_q$ on both sides of the glass-glass transition, vanishes. Such 
endpoint singularities have been described first for so-called schematic 
models of MCT where the microscopic details, i.e., the $q$-dependence, has 
been dropped in favor of mathematical simplicity. Rather than a precise 
interaction potential these models capture the mathematical structure of 
the problem. In these schematic models the transition points can be 
calculated analytically and the asymptotic expansions are not affected by 
limited numerical accuracy. For such schematic models, the asymptotic 
expansions for singularities $A_\ell$ with $\ell > 2$ have been performed: 
Two-step relaxation and time-temperature superposition become invalid, and 
-- most notably -- logarithmic decay laws emerge 
\cite{Goetze1989d,Goetze2002}.

The first microscopic model with an $A_3$~singularity in its parameter 
plane was found for Baxter's sticky hard-sphere model (SHSS) 
\cite{Baxter1968b}. The SHSS adds a short-ranged attraction to the 
hard-sphere potential similar to the square-well interaction, cf. Fig. 
\ref{fig:sss_pot}~(a). However, within the SHSS, depth $\Gamma$ and width 
$\delta$ can only be changed together while the product $\Gamma\delta$ 
remains fixed, in addition the limit $\delta\rightarrow 0$ is performed; 
this defines a so-called stickiness parameter $\tau$. Hence, the SHSS has 
two control parameters, packing fraction $\varphi$ like the HSS and the 
stickiness $\tau$. For the SHSS, the MCT predicts glass-glass transitions 
and an $A_3$~endpoint singularity \cite{Fabbian1999,Bergenholtz1999}. The 
glass-glass transitions take place between a repulsion-dominated glass 
state -- as known from the HSS -- and an attraction-dominated glass state 
that resembles bond formation as known from the gelation transition 
\cite{Bergenholtz1999}. While for a treatment of the SHSS within MCT, a 
cutoff wave-vector space needs to be introduced, it is by now well-known 
how this cutoff can be interpreted as an inverse length scale, and how the 
MCT results stay well-defined \cite{Goetze2003b}. In accordance with 
Whitney's theorem, for small changes in the cutoff neither fold nor cusp 
singularities change qualitatively.

\begin{figure}[htb]
\centerline{
\includegraphics[width=.45\columnwidth]{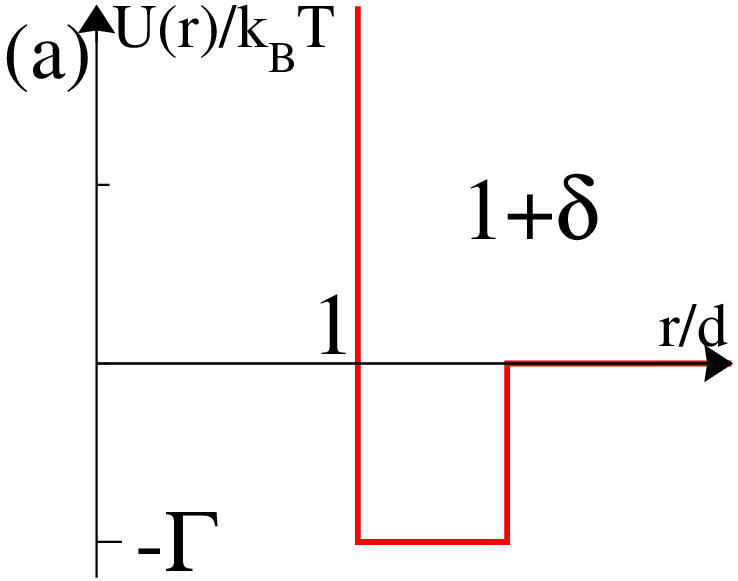}
\includegraphics[width=.45\columnwidth]{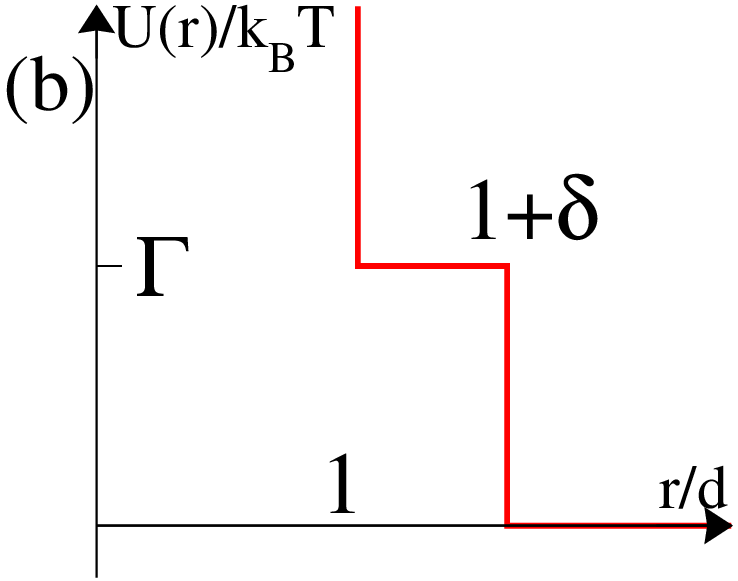}
}
\caption{\label{fig:sss_pot}(a) Square-well potential with three control 
parameters: packing fraction $\varphi = \pi\rho d^3/6$, well depth $\Gamma 
= u_0/k_\text{B}T$, and well width $\delta$ for particles of diameter $d$ 
at density $\rho$.
(b) Square-shoulder potential with shoulder height $\Gamma = 
u_0/k_\text{B}T$ and shoulder width $\delta$.
}
\end{figure}

\subsection{Glass-Glass Transitions in the Square-Well System -- a 
Swallowtail}\label{sec:SWS}

In order to avoid entirely the introduction of a cutoff and other 
peculiarities of the SHSS, one can extend the model attraction to finite 
widths in the square-well system (SWS), cf. Fig.~\ref{fig:sss_pot}~(a). 
Here, the control-parameter space becomes truly three dimensional, the 
parameter triple $(\varphi, \Gamma, \delta)$ defines each state. The MCT 
glass-transition scenarios have been worked out for the SWS 
\cite{Dawson2001}, and in addition to the cusp scenario of the SHSS there 
exists a characteristic well width $\delta^*$ for the SWS above which no 
cusp singularity can be found in the $(\varphi, \Gamma, \bar\delta)$ 
parameter plane for fixed $\bar\delta$ when $\bar\delta > \delta^*$. When 
$\bar\delta < \delta^*$, the SWS always exhibits a glass-glass transition 
line with an $A_3$~singularity as endpoint. For the exceptional point 
$\delta = \delta^*$, the glass-glass transition lines together with the 
line of $A_3$ singularities vanish in an $A_4$~singularity, giving rise to 
a three-dimensional geometric structure known as swallowtail. In contrast 
to the stable $A_2$ and $A_3$ singularities, the $A_4$ is an isolated 
point in the parameter space that is sensitive to small numerical 
deviations from $(\varphi^*, \Gamma^*, \delta^*)$. But while the details 
of the approximations involved and the numerical implementation change the 
location of the $A_4$~singularity, its existence is a robust prediction of 
MCT for systems with short-ranged attraction regardless of the specific 
interaction model \cite{Goetze2003b} or closure relation for the static 
structure factors \cite{Dawson2001}.

The occurrence of the glass-glass transitions can be traced to two 
different mechanisms of arrest -- resulting from the interplay of 
repulsion and attraction -- in the vertex, cf. Eq.(\ref{eq:eom:V}). The 
repulsion-dominated glass state within MCT is produced by the local 
structure on the wave-vector scale of the peak of $S_q$. The second 
mechanism of arrest is given by a $1/q$-tail in $S_q$ for large wave 
vectors $q$ \cite{Dawson2001}. Once that tail has enough weight in the 
vertex, a second transition -- the glass-glass transition -- can occur. 
This possibility of an additional transition is robust regarding the 
closure relation as long as the tail can become large enough. This 
independence of the results from the closure relation was demonstrated 
explicitly for the PY compared with the mean-spherical approximation 
\cite{Dawson2001,Sperl2004}.

The asymptotic solutions have been worked out in detail for the SWS and 
involve logarithmic decay laws \cite{Sperl2003a}, Vogel-Fulcher-like 
divergence of time-scales \cite{Sperl2004b}, unconventional (i.e., 
non-power-law) critical decays at the higher-order singularities 
\cite{Goetze2004c}, and the interplay of two $A_2$ singularities at the 
crossing of glass- and gel-transition lines \cite{Sperl2004}. Experimental 
verifications of the scenarios predicted for the SWS are found in computer 
simulations and in colloidal suspensions with attraction among the 
particles. Confirmations include the reentrant behavior of the lines of 
$A_2$~singularities \cite{Eckert2002,Foffi2002b,Pham2002}, the dynamics at 
a crossing of glass- and gel-transition lines for micellar 
\cite{Chen2002,Chen2003b} and colloidal suspensions \cite{Pham2004}, and 
the logarithmic decay of correlation functions together with novel power 
laws for the MSD \cite{Sciortino2003,Sperl2003a}.

\subsection{Glass-Glass Transitions in the Square-Shoulder System}
\label{sec:SSS}

When the attractive well of the SWS is replaced by a repulsive step, one 
obtains the square-shoulder system (SSS), cf. Fig.~\ref{fig:sss_pot}~(b). 
The SSS also has three control parameters, $(\varphi, \Gamma, \delta)$, 
with $\Gamma$ now symbolizing the height of a shoulder. Allowing the 
control parameter $\Gamma$ to carry a sign, SWS and SSS may even be 
plotted into a single diagram. Recently, the MCT predictions for the SSS 
have been worked out using the RY approximation \cite{Rogers1984} for 
$S_q$ \cite{Sperl2009}. Further details and references to the numerical 
algorithms can be found in recent work \cite{Sperl2009}.

\begin{figure}[htb]
\centerline{
\includegraphics[width=.67\columnwidth]{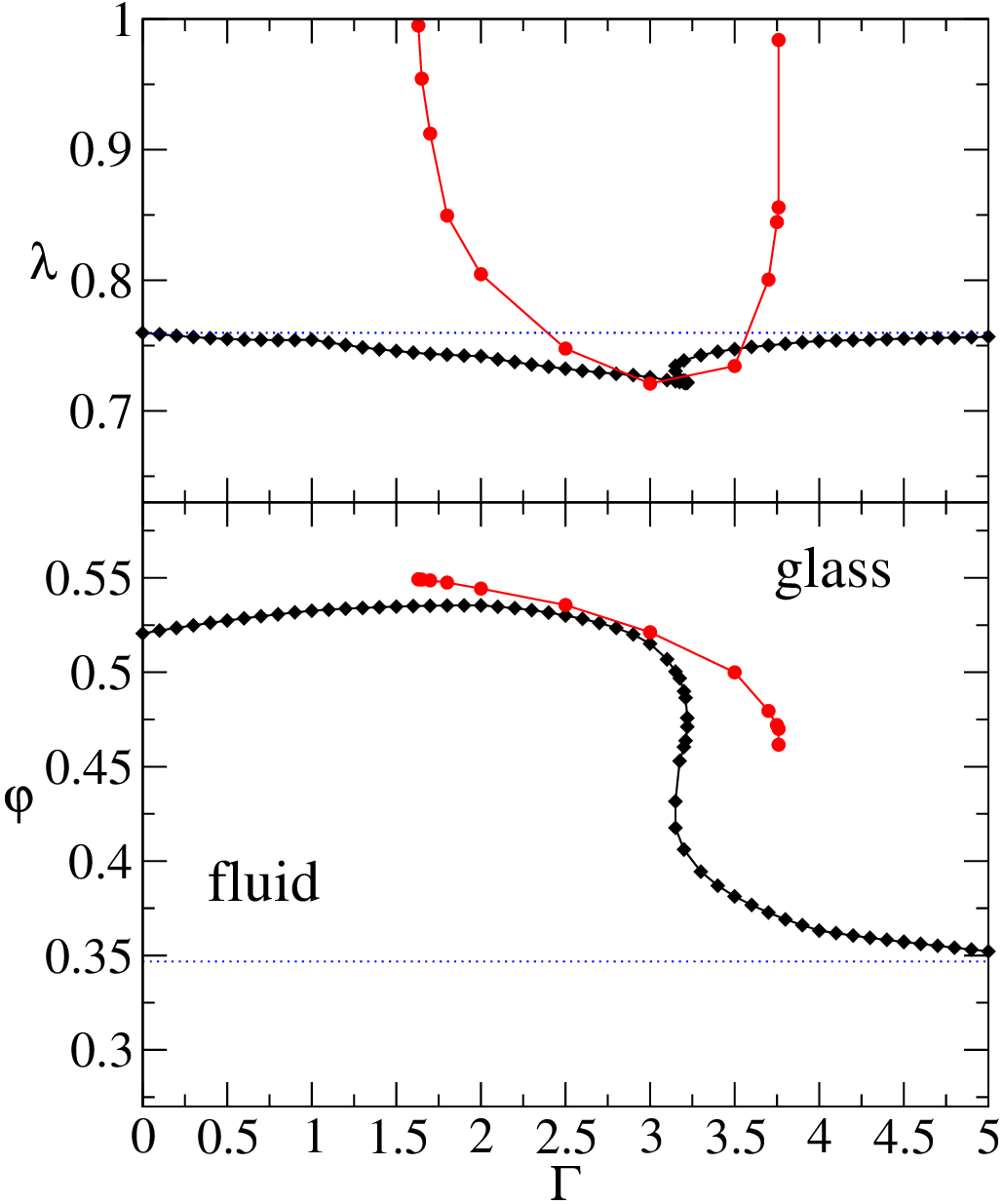}
}
\caption{\label{fig:sssry}Glass-transition diagram (lower panel) and 
exponent parameter $\lambda$ for the SSS using the RY approximation
at $\delta = 0.145$. Diamonds show the transition points from an ergodic 
state to a glass state; filled circles exhibit the glass-glass transition 
points. Dotted lines show the limits for the hard-sphere system with 
diameter $1+\delta$.
}
\end{figure}

Figure~\ref{fig:sssry} shows the results of MCT for the SSS at $\delta = 
0.145$: For $\Gamma = 0$, the glass transition line (lower panel) emerges 
from the HSS value of $\varphi_\text{HSS}^\text{RY} = 0.5206$, increases 
in density until around $\Gamma = 3$, bends over in an S-shape towards the 
HSS limit for the outer core, $\varphi = 0.5206/(1+0.145)^3 = 0.3468$. The 
form of the transition curve can be understood in detail from the 
distortions of the local structures by the presence of the repulsion at 
distance $1+\delta$ and the resulting changes to the principal peak of the 
static structure factor \cite{Sperl2009}. For both reentrant transitions 
it can be shown how the weakening of the local structure is compensated by 
either higher density (in the case of melting by cooling, i.e., melting at 
increasing $\Gamma$) or lower temperature which is equivalent to higher 
$\Gamma$ (in the case of melting by compression). In the 
pressure-temperature representation, such a reentrant behavior can be 
identified with a so-called diffusion anomaly which is known 
experimentally e.g. for water \cite{Angell1976}. The exponent parameter 
$\lambda$ (upper panel) varies very little along the glass-transition line 
and stays around the common value of $\lambda = 0.7$. From this 
unremarkable value for $\lambda$ no conclusion can be drawn about any 
glass-glass transitions or higher-order singularities in the vicinity.

Moving further into the glass state, $f_q$ experiences an additional jump 
marked by the full circles in Fig.~\ref{fig:sssry}. In contrast to the 
situation for the SWS, for the SSS the glass-glass-transition line is 
located completely within the glassy state. This gives rise to two 
endpoints, two $A_3$ singularities. These two endpoints are seen also in 
the upper panel of Fig.~\ref{fig:sssry} where $\lambda$ approaches unity 
on either side. This novel line originates from the competition of the two 
repulsive cores that causes a beating in the static structure factor for 
large wave vectors; this beating -- if sufficiently large in amplitude -- 
can bring additional weight to the MCT vertex in Eq.~(\ref{eq:eom:V}) and 
hence cause an additional discontinuous transition \cite{Sperl2009}. In 
physical terms, the localization of the already arrested particles drops 
drastically when the glass-glass transition line is crossed since upon 
crossing the line, the localization now happens at the outer core rather 
than at the inner core as before. When varying the control parameters, the 
beating is most pronounced when inner and outer core have about the same 
impact on $S_q$ -- if either one of the cores is dominant, the beating is 
diminished and the discontinuous transition ends in two $A_3$ 
singularities, respectively. Physically, the glass-glass transition 
becomes impossible at the upper endpoint when the density becomes too high 
for a transition to the outer core; the glass-glass transition also 
vanishes at the lower endpoint because the density becomes too low at the 
respective shoulder height to force the particles closer together. For 
shoulder widths $\delta$ larger than the value shown in 
Fig.~\ref{fig:sssry}, the glass-glass transition line moves towards and 
merges with the glass-transition line \cite{Sperl2009}.

\section{Square-Shoulder System using the Percus-Yevick Approximation}
\label{sec:SSSPY}

In this section, it is demonstrated how the scenario shown in 
sec.~\ref{sec:SSS} changes when the mentioned beating has not enough 
weight in the vertex~(\ref{eq:eom:V}) to cause an additional transition 
line.

\begin{figure}[htb]
\centerline{\includegraphics[width=.67\columnwidth]{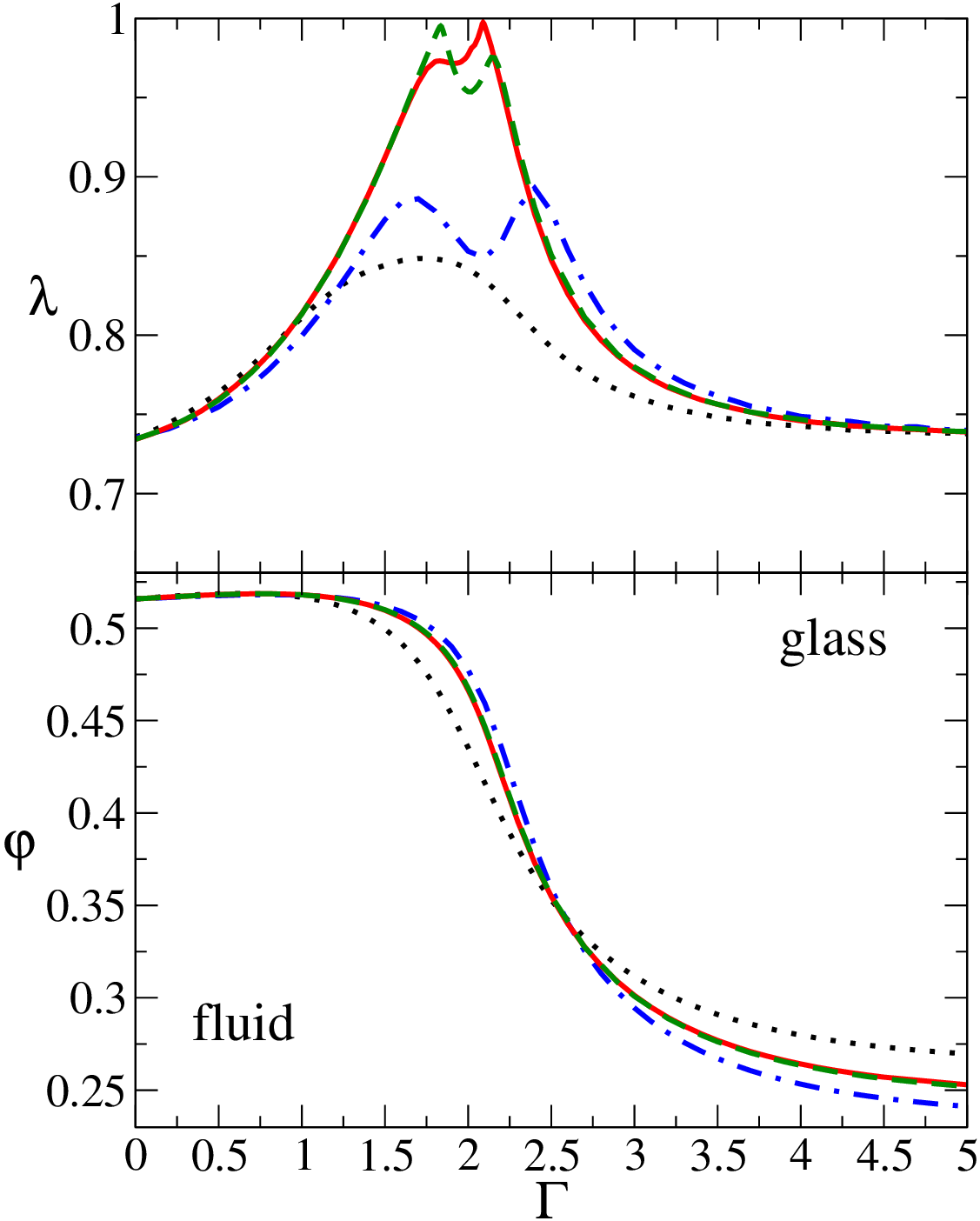}}
\caption{\label{fig:PDPYall}Glass-transition scenario (transition lines in 
the lower and exponent parameter $\lambda$ in the upper one) for the SSS 
using  the PY approximation for $\delta = 0.25$ (dotted line), 0.2785 
(full line), 0.28 (dashed line), and 0.3 (chain line).
In the lower panel, the curves for $\delta = 0.2785$ and 0.28 are almost 
on top of each other.
}
\end{figure}

Figure~\ref{fig:PDPYall} shows the glass-transition scenarios for various 
shoulder widths of the SSS when the PY approximation is used for the 
calculation of the static structure factor. The overall behavior of the 
transition diagram is similar to the RY results: The glass-transition 
first increases in packing fraction $\varphi$ when starting from the HSS 
limit $\varphi_\text{HSS} = 0.516$ at $\Gamma = 0$. Around $\Gamma = 2.5$ 
the curve bends downwards and reaches the limit of the HSS with an outer 
core of $\varphi_\text{HSS}/ (1+\delta)^3$ at around $\Gamma = 5$. 
Different from the RY result, the S-shape of the transition curve for 
intermediate values of $\Gamma$ only develops at higher values for the 
width $\delta$. Also in contrast to the RY result, there appears to be no 
indication of additional discontinuities in the $f_q$ or endpoint 
singularities inside the glass regime. However, while $\lambda$ in the 
upper panel of Fig.~\ref{fig:sssry} stays almost constant at the 
glass-transition line for the RY results, for the glass-transition line 
within PY approximation the exponent parameter $\lambda$ varies 
considerably with $\Gamma$ for any given width $\delta$ in the upper panel 
of Fig.~\ref{fig:PDPYall}. In addition, the shape of the 
$\lambda$-versus-$\Gamma$ curves varies drastically with $\delta$. For 
$\delta = 0.25$, an exponent-parameter maximum is around $\lambda = 0.85$ 
which is already rather high. For larger $\delta$, the 
$\lambda$-versus-$\Gamma$ curves exhibit double maxima over relatively 
small intervals in $\Gamma$, like between 1.85 and 2.1 for $\delta$ 
around 0.28. For $\delta = 0.2785$ and $\delta = 0.28$ the parameter 
$\lambda$ approaches unity very closely at the right and the left end of 
the interval, respectively. Beyond that regime, $\lambda$ decreases again 
for larger $\delta$, and as shown for $\delta = 0.3$, the separation of 
both $\lambda$ maxima increases; the $\lambda$ maxima are located at 
$\Gamma = 1.65$ and $\Gamma = 2.4$, respectively, while their values are 
still as high as $0.9$. In summary, while no glass-glass line can be 
detected, the exponent parameter approaches $\lambda = 1$ very closely. 
These results indicate a nearby higher-order singularity without the 
presence of glass-glass transitions. More puzzling still is the existence 
of two maxima in $\lambda$ very close to each other.

\begin{figure}[htb] 
\centerline{\includegraphics[width=.67\columnwidth]{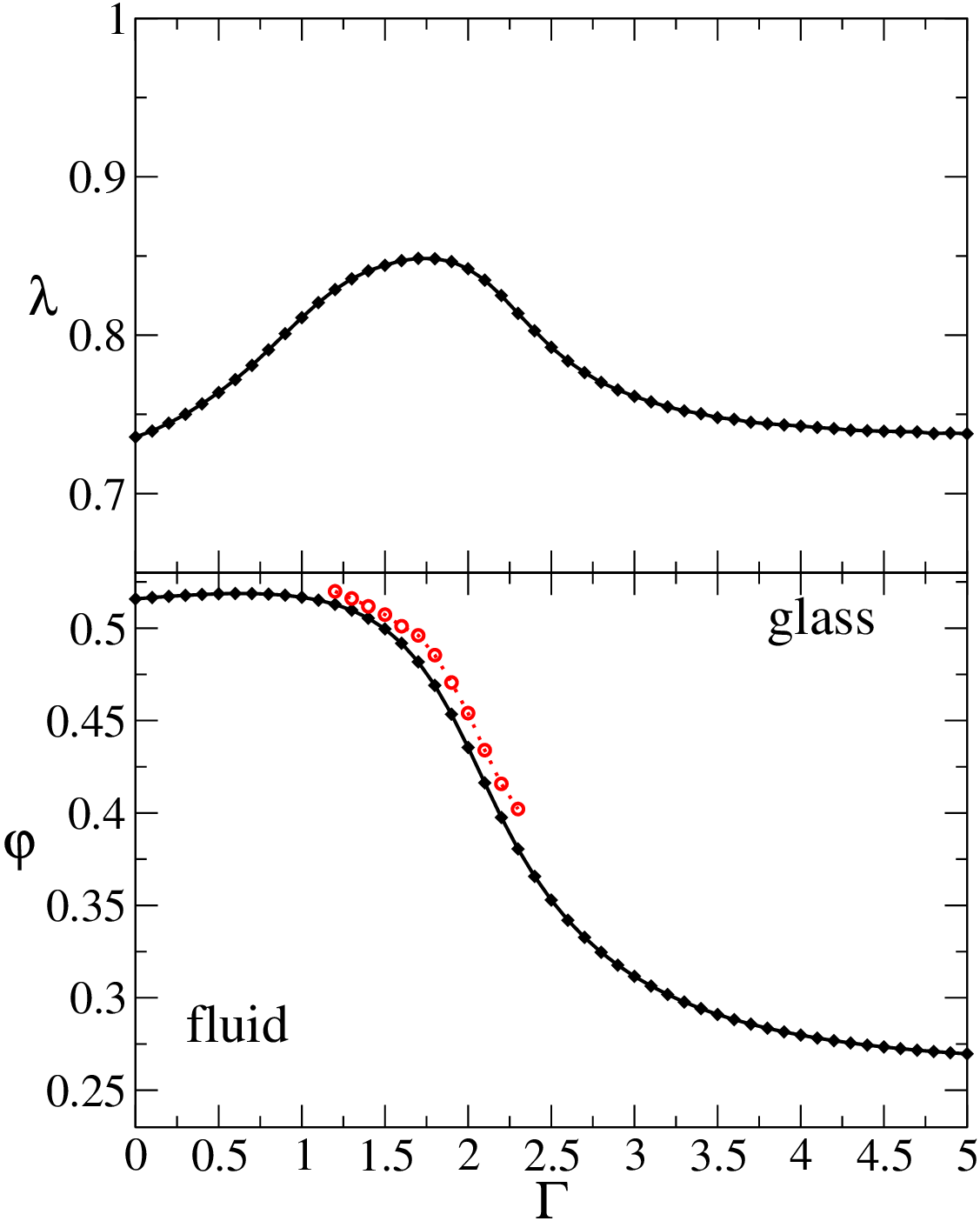}} 
\caption{\label{fig:PDPY25}Glass-transition scenario for the SSS for 
$\delta = 0.25$. Diamonds show the $A_2$ singularities at glass-transition 
points. The open circles mark the position of an anomaly in the evolution 
of the $f_q$, see text -- a line of hidden glass-glass transition points 
emerges.
} 
\end{figure}

To clarify the nature of the higher-order singularities on the 
glass-transition curves, the glassy region is inspected in more detail for 
$\delta = 0.25$ in Fig.~\ref{fig:PDPY25}. The behavior of $\lambda$ 
indicates that here the higher-order singularities are further away from 
the transition line, but a rather broad maximum already hints at these. It 
is well-known that after crossing an $A_2$ glass-transition singularity, 
the long-time limits $f_q$ increase above their critical value $f_q^c$ in 
a square-root in the control parameter; i.e. in the HSS, this increase is 
proportional to $\sqrt{\varphi-\varphi^c_\text{HSS}}$ \cite{Franosch1997}. 
When a glass-glass transition occurs, this square-root increase after the 
first transition is superseded by an additional discontinuity in $f_q$ 
followed by the square-root increase after this second transition. While 
this additional discontinuity was used to identify the glass-glass 
transition line in Fig.~\ref{fig:sssry}, such discontinuity is absent for 
the PY calculations. Nevertheless, the evolution of the $f_q$ within PY 
approximation shows a characteristic deviation from the square-root 
behavior in certain parameter regions: First, the range of validity of the 
square-root increase is sometimes far smaller than known from the HSS; 
second, the apparent square-root increase at larger distances from the 
glass-transition indicates a square-root with a different extrapolated 
transition point $\varphi^\text{app}$ than the one given by the 
discontinuity in $f_q$ at $\varphi^c$. This difference between the actual 
and the apparent transition point can be quantified as a difference in the 
control parameters, say as a relative difference in packing fraction, 
$\Delta\varphi = (\varphi^\text{app}-\varphi^c)/\varphi^c$. When this 
relative difference in packing fraction exceeds 1\%, this anomaly is 
marked by open circles in Fig.~\ref{fig:PDPY25}. It is seen that this line 
of $f_q$ anomalies strongly resembles the line of glass-glass transitions 
in Fig.~\ref{fig:sssry}. It can therefore be concluded that within MCT 
both approximations, PY and RY, yield similar glass-transition scenarios 
with a possible line of glass-glass transitions inside the glassy regime.

\section{Conclusion}

In the present work, the glass-transition diagram for the SSS has been 
calculated for the PY approximation. These results can now be used to 
estimate the robustness of the results for the SSS obtained with a 
different closure relation, e.g. with results from the RY approximation 
\cite{Sperl2009}. While the comparison of different closure relations for 
the SWS only resulted in shifts of the control parameters, the situation 
is more involved in the case of the SSS. The existence of the disconnected 
glass-glass transition line depends on two trends that both become more 
prominent for larger values of $\delta$. First, the vertex in 
Eq.~(\ref{eq:eom:V}) obtains additional weight for higher wave vectors 
through a beating phenomenon as described above; this makes a glass-glass 
transition possible if the weight becomes strong enough. Second, at the 
same time this supposed glass-glass transition line moves closer to the 
glass transition line and merges with it. For the RY approximation, the 
glass-glass transition line becomes manifest through a discontinuity in 
$f_q$ and then moves towards the glass transition line. In contrast for 
the PY approximation, the not-yet-manifest glass-glass transition moves 
towards the glass transition line and merges with it before it is fully 
developed as a discontinuity in the $f_q$. When the endpoints of this 
hidden line of glass-glass transitions merge with the regular glass 
transition line, very high values of $\lambda$ result since the first 
trend of increased weight for a glass-glass transition keeps increasing. 
In this sense, the glass-glass transition line emerges extremely close and 
on top of the glass transition line and is at the same moment absorbed by 
the glass transition line.

The differences between RY and PY approximation with respect to the 
higher-order singularities and the corresponding glass-glass transitions 
are highly non-trivial in the theoretical calculations. However, for the 
experimental test if such a scenario exists, both scenarios need to be 
observed in combination: Numerical deviations similar to the ones shifting 
the HSS-PY transition from $\varphi^c = 0.516$ to 0.58 in the experiment 
can for the SSS switch from the RY scenario shown earlier \cite{Sperl2009} 
to the PY scenario described here.

\section*{Acknowledgments}

This work was partially supported by Yukawa International Program for 
Quark-Hadron Sciences (YIPQS). Support from BMWi under 50WM0741 is 
gratefully acknowledged. I want to thank W.~G\"otze, J.~Horbach, P.~Kumar, 
F.~Sciortino H.~E.~Stanley and E.~Zaccarelli for fruitful collaboration 
and discussion of the work.

\end{document}